\documentclass[12pt]{iopart}

\usepackage{iopams}  
\usepackage{graphicx}
\usepackage{stix}
\usepackage{xcolor}
\usepackage{upgreek}
\begin{document}

\title[First-passage time distribution of a Brownian particle confined in a viscoelastic bath]{First-passage time distribution of a Brownian particle harmonically confined in a viscoelastic bath}

\author{Brandon R. Ferrer$^{1}$, Juan Ruben Gomez-Solano$^{1,*}$}

\address{$^{1}$Instituto de F\'isica, Universidad Nacional Aut\'onoma de M\'exico, Ciudad de M\'exico, C\'odigo Postal 04510, Mexico,}

\ead{$^*$ r\_gomez@fisica.unam.mx}
\vspace{10pt}
\begin{indented}
\item[]May 2026
\end{indented}

\begin{abstract}
We investigate theoretically and experimentally the first passage-time properties of a spherical Brownian particle that is harmonically trapped at thermal equilibrium in a fluid at constant temperature. By using the overdamped version of the generalized Langevin equation, we derive a general expression for the probability density function of the time that the particle takes to reach for the first time the minimum of the potential starting from an arbitrary position. We show that such a first-passage time distribution can be implicitly expressed in terms of the friction memory kernel that encodes the interaction of the particle with its surroundings, and correctly reduces to previously found expressions in the case of a Markovian viscous bath.  We validate our theoretical results by measuring the first-passage time of colloidal beads optically trapped in non-Markovian baths such as viscoelastic polymer and micellar solutions, as well as in a viscous glycerol/water mixture and water, which behave as Markovian media, having quantitative agreement with the derived expressions. In particular, we find that the mean first-passage time in a viscoelastic bath can surpass that in a viscous medium of the same zero-shear viscosity due to the emergence of slowly decaying tails in the first-passage time probability density of the former. 
\end{abstract}

%
\vspace{2pc}

%
%
%

\section{Introduction}\label{sect:intro}

The study of stochastic search processes has been a subject of major importance in statistical mechanics due to its ubiquity in nature. In a random search problem, it is useful to determine the time it takes a system subject to noise to reach a target for the first time starting from a given initial condition~\cite{metzler2014}. This time is a stochastic variable whose statistical properties are of prime importance in a wide variety of processes, such as diffusion controlled reactions~\cite{szabo1980}, protein binding to DNA \cite{ghosh2018}, single-molecule experiments~\cite{dudko2008}, intracellular transport~\cite{mogre2020},
animal foraging~\cite{viswanathan2011}, epidemic spread \cite{kumari2024}, and decision-making processes \cite{broderick2010}, among others. Its expected value, which is called the mean first-passage time (MFPT), provides information on the efficiency of the search process, and depending on the particular strategy, it may be finite or divergent as in the case of a free Brownian motion, in which case a
purely diffusive strategy is completely inefficient \cite{chandrasekhar1943,gupta2022}. In the context of random-walk processes, numerous efforts have been made to theoretically analyze first-passage time properties of Brownian systems under various conditions, such as intermittent search strategies~\cite{benichou2005,benichou2006,oshanin2007,lomholt2008,mercadovasquez2019}, barrier-crossing dynamics \cite{reimann1999,zhou2014,berezhkovskii2019,abdoli2023,ferrer2024}, diffusion under geometrical confinement \cite{benichou2014,antoine2016,grebenkov2017,berezhkovskii2017,perezespinosa2020,pompagarcia2022}, dynamics in confining potentials \cite{hu2010,grebenkov2014,walter2021,mercadovazquez2022,huang2025}, nonlinear diffusion~\cite{wang2008,wang_2008,dostal2020,chelminiak2024}, searchers with anisotropic shape ~\cite{levernier2017,gosh2020,gosh2022,gosh2026}, motion of active particles~\cite{angelani2014,scacchi2018,bressloff2023,iyaniwura2024,baouche2025,iyaniwura2025}, and multiple-particle systems \cite{
wang08,basnayake2018,grebenkov2020,hass2024}. Resetting the system repeatedly at periodic or random times to an initial state can minimize the MFPT to the target of many of the aforementioned processes at certain optimal rates, representing a type of search strategy that has attracted a great deal of attention in recent years ~\cite{gupta2022,gosh2026,evans2011,nagar2016,pal2019,gupta2019,
evans2020,singh2020,santra2021,abdoli2021,jain2023,villanuevaalcala2024,baouche_2025,chaki2025,najeeb2026}.

It should be noted that most of the investigations on first-passage-time properties of stochastic search processes assume Markovianity, since it is a good  approximation for many systems immersed in simple environments. Nevertheless, non-Markovian dynamics resulting from the interaction of the system with slow degrees of freedom of the surroundings are widespread in many situations, such as tagged monomers in long polymer chains \cite{sakaue2013,panja2010,durang2024}, mesoscopic objects undergoing single-file diffusion in narrow channels~\cite{lutz2004,eab2010,grabsch2024}, or subject to hydrodynamic backflow in liquids~\cite{kheifets2014,gomezsolano2024,boynewicz2026}, colloids passively moving in viscoelastic fluids \cite{grebenkov_2014,darabi2023,ferrer_2024}, and probe particles embedded in non-equilibrium active baths \cite{guevaravaladez2023,goswami2023,boudet2025}. In such instances, the current state of the system generally depends on its past history, and therefore memory effects cannot be neglected in the description of its first-passage statistics \cite{bray2013}, which has been the subject of a number of investigations over the past years \cite{benichou2015}. Early examples of these include the calculation of first-passage time distributions of a tagged monomer in a polymer chain \cite{amitai2010}, for different models of subdiffusion~\cite{koszto2012}, tracers flanked by identical crowders displaying single-file diffusion
\cite{sanders2012}, or with additional binding/unbinding dynamics~\cite{forsling2014}, and Brownian particles moving in viscoelastic fluids with a Rouse friction kernel \cite{hohenegger2017}. Further theoretical developments enabled the derivation of general analytical expressions for the MFPT of non-Markovian fractional Brownian walkers with stationary increments in terms of their mean square displacement and their average trajectory after the first-passage event in the case of geometrical confinement \cite{guerin2016} and for rarely visited configurations \cite{levernier2020}. The same approach could be applied to the study of the persistence probability of non-Markovian systems with non-stationary dynamics relaxing to a steady state after an initial perturbation \cite{levernier2022}. Moreover, the statistical properties of first-passage times of subdiffusive random walkers were successfully  addressed using an alternative dynamical response approach~\cite{sakamoto2023}. Memory effects have also been theoretically and experimentally quantified through the calculation of splitting probabilities in competitive first-passage events of micrometer-sized beads in viscoelastic polymer solutions \cite{dolgushev2025}.
All of these works have triggered the recent analysis of the impact of stochastic resetting protocols on the first-passage time statistics of Brownian particles in viscoelastic baths,  both theoretically \cite{goswami2025,biswas2025,jolakoski2025,mondal2025} and experimentally \cite{ginot2026}, revealing the emergence of novel strategies of optimal search in non-Markovian environments due to the interplay of the externally imposed resetting rate with the intrinsic memory time-scales of the system.

In this paper, we investigate the first-passage time statistics of a Brownian system of broad significance in the context of soft matter physics: a colloidal bead in thermal equilibrium with a fluid medium at constant temperature, whose motion is confined by a harmonic potential. Our analysis is not restricted to a specific shape of the friction memory kernel of the dynamics, thus allowing us to explore the cases of viscous media theoretically studied in the past, as well as viscoelastic fluids with arbitrary relaxation moduli like those observed in experiments. In particular, we derive an expression for the probability density function of the first-passage time of the particle to reach the potential minimum starting from an arbitrary location, which can be explicitly expressed in terms of a time-dependent function directly related to the stationary autocorrelation function of its position, whose derivative dictates the asymptotic long-time behavior. These results are experimentally validated by measuring the first-passage time distributions of micron-sized spheres trapped by optical tweezers in viscous media like water and a glycerol/water mixture and in two types of viscoelastic fluids: an aqueous polymer solution and a micellar solution, where the two latter cases exhibit slowly decaying tails that give rise to values of the MFPT that are larger than those observed in a viscous fluid if the particle initially starts with a sufficiently large potential energy.

The paper is organized as follows. In Section \ref{sect:model} we present the model for the motion of a Brownian particle confined by a harmonic potential in a viscoelastic fluid at constant temperature. In Section \ref{sect:FPT_distribution} we derive an expression of the corresponding first-passage time distribution, and discuss its main features and its relation to previous formulae derived for particles moving in viscous fluids. The experimental details are provided in Section \ref{sect:exp}, while the comparison of the experimental results of the first-passage time statistics with the theoretically derived expressions and the corresponding discussion are reported in Section \ref{sect:res}. Finally, we summarize our main results and conclude in Section \ref{sect:summ}.

\section{Model}
\label{sect:model}

We are interested in the first-passage time properties of a micrometric spherical bead of radius $a$ harmonically trapped in a homogeneous and isotropic fluid kept at constant temperature $T$. The linear rheological properties of the fluid are quantified by its relaxation modulus, $G(t)$, whose dependence on time $t$ satisfies $G(t \rightarrow \infty) = 0$ and $G(t) = 0$ for $t < 0$ by causality \cite{bird1987dynamics}. For the sake of simplicity, we focus on the overdamped dynamics of a single coordinate of the center of mass of the bead, $x$, which is described as by the generalized Langevin equation~\cite{kubo1966}
\begin{equation}\label{eq:GLE}
    0 = -\int_0^t dt' \, \Gamma(t-t') \dot{x}(t') -\kappa x(t) + \zeta(t).
\end{equation}
In Eq.~(\ref{eq:GLE}), inertial effects are neglected on the left-hand side, which is a good approximation for the motion of colloidal particles moving in liquids on time-scales much longer than their momentum relaxation time and the vortex diffusion time due to hydrodynamic backflow. 
Furthermore, the first term on the right-hand side of~Eq. (\ref{eq:GLE}) represents the drag force acting on the bead at time $t$, where $\Gamma(t-t')$ is a memory kernel that quantifies the delayed effect of the fluid at times $0 \le t' \le t$, with $\Gamma(t - t') = 0$ for $t' > t$ due to causality. In this overdamped limit, the Laplace transform of the memory kernel, $\tilde{\Gamma}(s) = \int_0^{\infty} dt\, e^{-st} \Gamma(t)$, is linked to the Laplace transform of the fluid's relaxation modulus, $\tilde{\eta}(s) \equiv \tilde{G}(s)  = \int_0^{\infty} dt \, e^{-st} G(t)$, through the Stokes relation $\tilde{\Gamma}(s) = 6\pi a \tilde{\eta}(s)$, where $\tilde{\eta}(s)$ represents a complex viscosity dependent on the Laplace frequency, $s$, thus accounting for the linear viscoelastic behavior of the fluid. Moreover, $ -\kappa x(t)$ is a restoring force exerted on the particle at time $t$ by a harmonic potential of stiffness $\kappa$, $U(x) = \frac{1}{2} \kappa x^2$. In addition, $\zeta(t)$ is a Gaussian colored noise that results from the thermal fluctuations of the fluid particles, whose mean and autocorrelation function at thermal equilibrium are given by
\begin{eqnarray}\label{eq:FD2nd}
    \langle \zeta(t) \rangle & = & 0, \nonumber\\
    \langle \zeta(t) \zeta(0) \rangle & = & k_B T \Gamma(t),
\end{eqnarray}
respectively. 

The solution of Eq. (\ref{eq:GLE}) for a particle trajectory starting from the initial condition $x(0)$ can be expressed as
\begin{equation}\label{eq:solx}
    x(t) = x(0) \chi(t) + \int_0^{t} dt' \, \varsigma(t-t') \zeta(t'),
\end{equation}
where $\chi(t)$ and $\varsigma(t)$ are the inverse Laplace transforms of the functions $\tilde{\chi}(s)$ and $\tilde{\varsigma}(s)$ defined by the equations
\begin{equation}\label{eq:chi}
    \tilde{\chi}(s) = \frac{\tilde{\Gamma}(s)}{ s \tilde{\Gamma}(s) + \kappa},
\end{equation}
and
\begin{equation}\label{eq:sigma}
    \tilde{\varsigma}(s) = \frac{1}{s\tilde{\Gamma}(s) + \kappa},
\end{equation}
respectively. Irrespective of the specific form of the memory kernel, the function $\chi(t)$ given by the inverse Laplace transform of Eq. (\ref{eq:chi}), has the following asymptotic values
\begin{eqnarray}\label{eq:chiasympt}
    \lim_{t \rightarrow 0} \chi(t) & = \lim_{s \rightarrow \infty} s \tilde{\chi}(s) & = 1, \nonumber\\
    \lim_{t \rightarrow \infty} \chi(t) & = \lim_{s \rightarrow 0} s \tilde{\chi}(s) & = 0,
\end{eqnarray}
where it has been taken into account that $\lim_{s \rightarrow 0} s\tilde{\Gamma}(s) = \lim_{t \rightarrow \infty } \Gamma(t) = 0$ since $G(t \rightarrow \infty) = 0$ for a fluid.

As demonstrated in \cite{okuyama1986,chaudhury2006}, for a particle with non-Markovian dynamics described by Eq. (\ref{eq:GLE}), the probability density of finding it in position $x$ at time $t > 0$ provided that it was located at $x_0 = x(0)$ at time $t = 0$ with open boundary conditions, can be expressed as
\begin{equation}\label{eq:condpdf0}
    P_0(x,t|x_0,0) = \sqrt{\frac{\kappa}{2\pi k_B T \left[ 1 - \chi(t)^2\right]}} \exp \left\{ - \frac{\kappa \left[ x  - x_0 \chi(t)\right]^2}{2k_B T \left[ 1 - \chi(t)^2\right]}\right\}.
\end{equation}
Note that if one assumes that at time $t=0$ the particle is already in thermal equilibrium, with $x_0$ drawn from the canonical distribution $P_{eq}(x_0) = \sqrt{\frac{\kappa}{2\pi k_B T}} e^{-\frac{\kappa x_0^2}{2 k_B T}}$, then it remains so for all times $t > 0$, i.e. $P_{eq}(x) = \int_{-\infty}^{\infty} dx_0 P(x,t|x_0,0) P_{eq}(x_0)$, where the stationary  autocorrelation function of its position is $\langle x(\tau) x(0) \rangle = \langle x(t + \tau)x(t) \rangle = \frac{k_B T}{\kappa} \chi(\tau)$ \cite{okuyama1986}. This provides a physical interpretation for the function $\chi(t)$, either as the dimensionless stationary auto-correlation function of the particle position at thermal equilibrium or as the relaxation function from a given initial position. We point out that in the situation where $\tilde{\eta}(s) = \eta$ is independent of the frequency $s$, which corresponds to a Newtonian behavior of the fluid, the memory kernel becomes $\Gamma(t) = 2 \gamma \delta(t)$, in which case $\chi(t) = e^{-\kappa t/{\gamma}}$ and $\varsigma(t) = \gamma^{-1} e^{- \kappa t / {\gamma}}$ with $\gamma = 6\pi a \eta$ the constant friction coefficient experienced by the particle.

\section{First passage time distribution}
\label{sect:FPT_distribution}

We now proceed to derive an expression for the probability density function of the time $t_{\mathrm{FP}} >0$ taken by the Brownian particle to reach for the first time the minimum $x = 0$ of the harmonic potential $U(x) = \frac{1}{2} \kappa x^2$ when starting from an arbitrary position $x_0 \neq 0$ at $t=0$, which we denote as $f(x_0,t_{\mathrm{FP}})$. To find this first-passage time distribution, we follow the integral equation approach outlined in \cite{hu2010}, which is based on the calculation of the survival probability of finding the particle without having
reached at time $t>0$ the target located at $x = 0$, conditioned to be initially placed at $t=0$ in $x = x_0$. If $x_0 < 0$, the survival probability is
\begin{equation}\label{eq:surv}
    S(x_0,t) = \int_{-\infty}^0 dx P(x,t | x_0,0),
\end{equation}
where $P(x,t | x_0,0)$ is the conditional probability density at time $t > 0$ of the particle position, $x$, provided that it was in position $x_0$ at $t=0$, subject to an absorbing boundary condition at $x = 0$. Likewise, if $x_0 > 0$ the corresponding survival probability is 
\begin{equation}\label{eq:surv1}
    S(x_0,t) = \int_0^{\infty} dx P(x,t | x_0,0).
\end{equation}
In Eqs. (\ref{eq:surv}) and (\ref{eq:surv1}), $P(x,t | x_0,0)$ can be expressed in terms of the corresponding conditional probability density in the absence of the absorbing point $x = 0$, i.e. $P_0(x,t | x_0,0)$ given by Eq. (\ref{eq:pdf0}), and the probability density of the first-passage time through $x=0$, $f(x_0,t)$, by means of the following relation \cite{hu2010,guerin2016,sakamoto2023}
\begin{equation}\label{eq:pdf0}
    P_0(x,t|x_0,0) = P(x,t|x_0,0) + \int_0^t dt' f(x_0, t')   P_0(x,t|0,t').
\end{equation}
First, we analyze the case $x_0 < 0$, where it should be noted that on the right-hand side of Eq. (\ref{eq:pdf0}) the first term is a contribution to $P_0(x,t|x_0,0)$ of stochastic events that survived from $x_0$ to position $x < 0$ without having passed over the point $x=0$, while the second term represents those events in which the particle reached the target earlier at $t' < t$ and then returned to $x < 0$ at time $t$, where $P_0(x,t|0,t')$ is the corresponding probability density. Therefore, upon integration over the interval $-\infty < x < 0$, Eq. (\ref{eq:pdf0}) becomes
\begin{equation}\label{eq:intpdf0}
   \int_{-\infty}^0 dx P_0(x,t|x_0,0) = S(x_0,t) - \int_0^t dt' \frac{\partial S(x_0, t')} {\partial t'} \int_{-\infty}^0 dxP_0(x,t|0,t'),
\end{equation}
where we used Eq. (\ref{eq:surv}), and took into account that the survival probability at $t > 0$ is equal to the probability that the particle reaches the target $x=0$ for the first time only at a later instant $t' > t$,  i.e.  $S(x_0,t) = \int_t^{\infty} dt' f(x_0,t')$, or equivalently
\begin{equation}\label{eq:fpt}
    f(x_0,t_{\mathrm{FP}}) = - \left.\frac{\partial S(x_0,t)}{\partial t} \right|_{t = t_{\mathrm{FP}}}.
\end{equation}    
Additionally, the term $\int_{-\infty}^0 dxP_0(x,t|0,t')$ represents the probability that the particle, in the absence of the absorbing point, moves to a position $x < 0$ at time $t$ starting from the potential minimum at time $t' < t$. Due to the symmetry of the potential $U(x)$ with respect to $x = 0$, this probability is $1/2$, which implies that 
\begin{equation}\label{eq:intpdf1}
    \int_{-\infty}^{0} dx P_0(x,t|x_0,0) = \frac{1}{2} S(x_0,t) + \frac{1}{2},
\end{equation}
where we used the fact that $S(x_0,0) = 1$. Hence, using the explicit form of $P_0(x,t|x_0,0)$ expressed by Eq. (\ref{eq:condpdf0}), Eq. (\ref{eq:intpdf1})
yields the following expression for the survival probability $S(x_0,t)$ in the case $x_0<0$ for all $t > 0$
\begin{equation}\label{eq:survival}
    S(x_0,t) = \mathrm{erf} \left( \frac{|x_0| \chi(t)}{\sqrt{\frac{2 k_B T}{\kappa} \left[1 -\chi(t)^2 \right]} }\right),
\end{equation}
where $|x_0| = -x_0$. A similar calculation can be carried out for $x_0 > 0$, in which case the distinct terms of Eq. (\ref{eq:pdf0}) must be integrated over the interval $0 < x < \infty$, thereby leading to the same expression for the survival probability as in Eq. (\ref{eq:survival}) with $|x_0| = x_0$. Consequently,  using Eq. (\ref{eq:fpt}), for any initial position $x_0 \neq 0$ the corresponding distribution of the first-passage time $t_{\mathrm{FP}}$ is given by the following formula
\begin{equation}\label{eq:fptd}
    f(x_0,t_{\mathrm{FP}}) = - \frac{\sqrt{2 \kappa} | x_0 |}{\sqrt{ \pi k_B T \left[ 1 - \chi(t_{\mathrm{FP}})^2  \right]^3}} \frac{d \chi(t_{\mathrm{FP}})}{dt_{\mathrm{FP}}} \exp \left\{ - \frac{\kappa x_0^2 \chi(t_{\mathrm{FP}})^2}{2 k_B T \left[1 - \chi(t_{\mathrm{FP}})^2 \right]} \right\}. 
\end{equation}
We point out that Eq. (\ref{eq:fptd}) correctly reduces to a previously derived expression for the probability density of the first-passage time to $x=0$ of a Brownian particle moving under the action of a harmonic potential $U(x) = \frac{1}{2}\kappa x^2$ in a viscous medium with a constant friction coefficient $\gamma$ \cite{szabo1980,hu2010}, where $\chi(t) = e^{-\kappa t/\gamma}$, thus leading to
\begin{equation}\label{eq:fptd_newton}
    f(x_0,t_{\mathrm{FP}}) = \frac{\kappa}{\gamma}\frac{\sqrt{2 \kappa} | x_0 |}{ \sqrt{ \pi k_B T \left( 1 - e^{-{2\kappa t_{\mathrm{FP}}}/{\gamma}}  \right)^3}}  \exp \left[ -\frac{\kappa}{\gamma}t_{\mathrm{FP}} - \frac{\kappa x_0^2 e^{-2\kappa t_{\mathrm{FP}} / \gamma}}{2 k_B T \left(1 - e^{-2\kappa t_{\mathrm{FP}} / \gamma} \right)} \right]. 
\end{equation}

\section{Experimental description}\label{sect:exp}

\subsection{Experimental setup}\label{subsect:setup}

\begin{figure}
    \centering
\includegraphics[width=0.9\columnwidth]{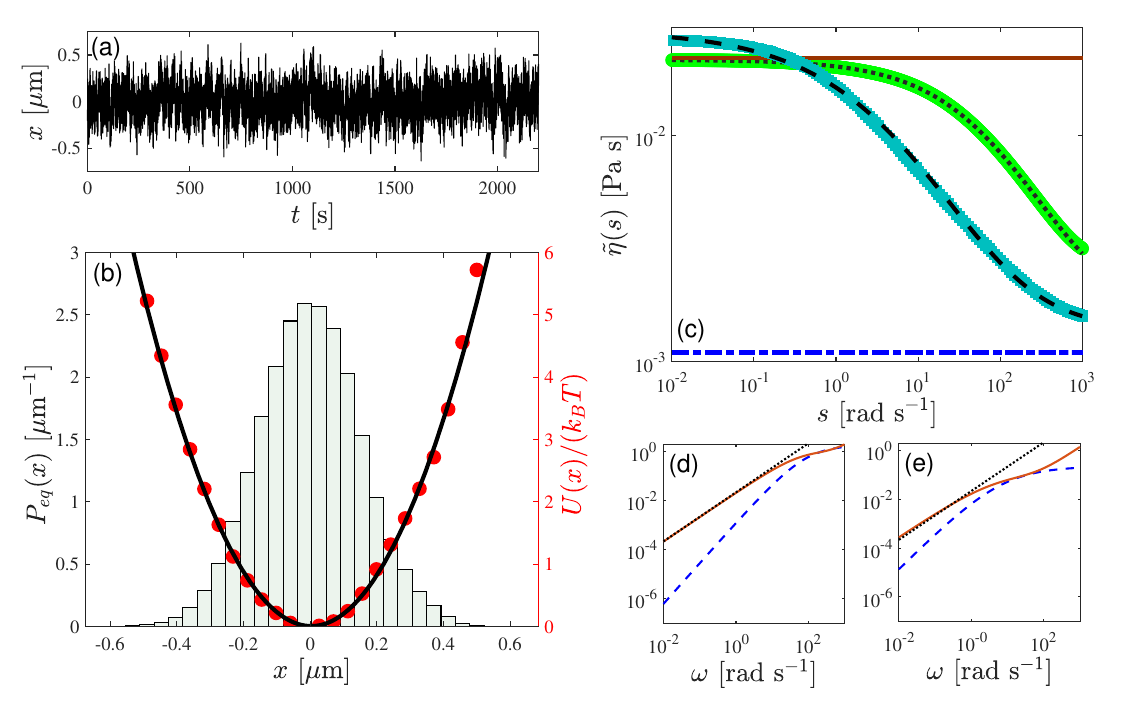}
    \caption{(a) Section of 40~min of the stochastic time evolution of the coordinate $x$ of the position of a spherical colloidal particle (radius $a = 1.0\,\mu$m) trapped in an aqueous micellar solution (temperature $T = 22^{\circ}$C) by optical tweezers at constant power ($P \approx 0.1$~mW). (b) Equilibrium probability density function of the coordinate $x$ of the same particle (vertical bars), experimental profile of the optical trapping potential ($\bigcirc$), and the corresponding quadratic fit (solid line), from which the stiffness of this harmonic trap is calculated: $\kappa=1.77\times10^{-7}~\mathrm{N~m}^{-1}$. (c) Viscosity as a function of the real Laplace frequency $s > 0$ of the polymer ($\bigcirc$) and the micellar solution ($\square$), where the dotted and dashed lines represent the curves using the fit function of Eq. (\ref{eq:etas}). The solid and dotted-dashed lines depict the frequency-independent viscosity of the glycerol/water mixture and the ultrapure water, respectively. 
    (d) Storage (dashed line) and loss modulus (solid line) of the polymer solution. (e) Storage (dashed line) and loss modulus (solid line) of the micellar solution. In (d) and (e), the Newtonian loss modulus of the glycerol/water mixture is traced as a dotted line as a reference.}\label{fig:1}
\end{figure}

To investigate experimentally the first-passage time properties of the Brownian system previously analyzed, we utilize an optical tweezers setup to create a harmonic potential acting on the motion of a colloidal bead immersed in a fluid. To this end, we dispersed spherical silica particles (radius $a = 1.0\,\mu$m) in the fluid of interest, which was hermetically confined in a sample cell made of a glass slide stuck to a coverslip with a double-sided adhesive tape (thickness $100\,\mu$m). A harmonic optical potential was generated by focusing a Gaussian laser beam (Opus 532, Laser Quantum, wavelength $\lambda = 532$~nm) inside the sample cell by an oil-immersion objective (Olympus Plan Fluorite $100\times$, NA~=~1.3), thus trapping a single colloidal bead at a distance of $h = 20\,\mu$m from the lower solid wall of the sample cell. Using a CMOS camera (Basler acA800-510um), we recorded
videos of the trapped particle at a sampling frequency of 1000 frames per second for 1 hour, from which the coordinates $(x,y)$
of the center of mass of the bead projected on the transverse plane to the beam direction were determined by standard particle-tracking algorithms with a spatial resolution of 10 nm. All experiments were performed at a temperature of $T = 22^{\circ}$C. Without loss of generality, we can focus on the stochastic evolution with time of a single coordinate, e.g. $x$, like the one shown in Fig. \ref{fig:1}(a). From the equilibrium probability density of the coordinate $x$, $P_{eq}(x)$, the trapping potential exerted by the optical tweezers on the particle was retrieved: $U(x) = -k_B T \ln \left[ P_{eq}(x) / P_{eq}(0) \right]$, which can be very well fitted to a quadratic function, $U(x) = \frac{1}{2}\kappa x^2$, as illustrated in Fig. \ref{fig:1}(b). The corresponding stiffness $\kappa$ was determined from the equilibrium variance $\langle x^2 \rangle$ of the particle position using the equipartition relation $\kappa = k_B T / \langle x^2 \rangle$. By selecting the laser power within the range $0.1 - 10$~mW, the values of $\kappa$ could be changed between $\sim 10^{-7}$~N~m$^{-1}$ and $\sim 10^{-6}$~N~m$^{-1}$, which enables examining the strength of confinement of potential in the statistics of the first-passage time of the bead in the different fluid media used in the experiments.

\subsection{Microrheological characterization of the fluids}\label{subsect:microrheo}

\begin{table}[ht!]
\centering 
\caption{Parameters characterizing the relaxation modulus of each fluid used in the optical-trapping experiments.}
\label{tab:1}
\begin{tabular}{ccccc}
\hline
Fluid & $\eta_0$ [Pa~s] & $\eta_\infty$ [Pa~s] & $\tau_0$ [s] & $\alpha$ \\ \hline
\textit{Ultrapure water} & $0.0011~\pm~0.0003$ & $-$ & $-$ & $-$ \\
\textit{Glycerol/water mixture} & $0.0220~\pm~0.0021$ & $-$ & $-$ & $-$ \\
\textit{Polymer solution} & $0.0216~\pm~0.0005$ & $0.0016~\pm~0.0002$ & $0.0105~\pm~0.0001$ & $0.658~\pm~0.002$ \\
\textit{Micellar solution} & $0.0292~\pm~0.0004$ & $0.0014~\pm~0.0003$ & $0.1220~\pm~0.0081$ & $0.490~\pm~0.012$ \\ \hline
\end{tabular}
\end{table}

To explore the effect of the medium on the stochastic motion of a harmonically confined particle, two Newtonian fluids 
and two non-Newtonian ones were used in the experiments as Markovian and non-Markovian baths, respectively. The first Newtonian fluid was simply ultrapure water (resistivity 18.2 M$\Omega$~cm at 25$^{\circ}$), which serves as the reference Newtonian case in this study, with a viscosity that does not depend on frequency and has a constant value of $\tilde{\eta}(s) = \eta_0  = 0.0011 \pm 0.0003$~Pa~s. The second Newtonian fluid was composed of glycerol at 70\%~wt mixed in ultrapure water, whose frequency-independent viscosity is higher than that of water and has the constant value $\eta(s) = \eta_0  = 0.0220 \pm 0.0021$~Pa~s. Moreover, the two non-Newtonian fluids were selected to exhibit viscoelasticity owing to their complex microstructure, i.e. a dependence on frequency of the complex viscosity $\tilde{\eta}(s)$. They were prepared by continuously stirring their corresponding components over 24 hours, thus becoming homogeneous and transparent, as required for the optical trapping experiments. One of them consisted of a polymer solution made of polyethylene oxide (molecular weight $4\times10^6$~Da) mixed in ultrapure water at 0.3\% wt.
The second viscoelastic fluid was an aqueous solution of elongated micelles composed of cetylpyridinium chloride and sodium salicylate at an equimolar concentration of 5 mM in ultrapure water. Using passive microrheology with optical tweezers \cite{darabi2023}, we determined the Laplace-frequency dependent viscosity of both fluids, which can be fitted by the function
\begin{equation}\label{eq:etas}
       \tilde{\eta}(s)=\eta_\infty+\frac{\eta_0-\eta_\infty}{\left[1+(\tau_0 s)^{\alpha}\right]^{1/\alpha}},
\end{equation}
whose inverse Laplace transform corresponds to the following form of the stress relaxation modulus
\begin{equation}\label{eq:Gt}
    G(t) = 2\eta_{\infty}\delta(t) + \frac{\eta_0 - \eta_{\infty}}{\tau_0} E_{\alpha,1}^{1/\alpha}\left(-\frac{t^{\alpha}}{\tau_0^{\alpha}} \right),
\end{equation}
where 
\begin{equation}\label{eq:MLfunc}
    E_{\alpha,\beta}^{\gamma} (z) = \sum_{k=0}^{\infty} \frac{\mathit{\Gamma}(\gamma + k)}{\mathit{\Gamma}(\gamma)\mathit{\Gamma}(\alpha k + \beta)} \frac{z^k}{k!},
\end{equation}
is the three-parameter Mittag-Leffler function, with $\mathit{\Gamma}(z) = \int_0^{\infty} du \, u^{z-1} e^{-u}$ the gamma function. In Eqs. (\ref{eq:etas}) and (\ref{eq:Gt}), $\eta_0 = \tilde{\eta}(s \rightarrow 0)$ and $\eta_{\infty} = \tilde{\eta}(s \rightarrow \infty)$ represent the zero-shear viscosity of the fluid and the viscosity of its solvent, respectively, $\tau_0$
is its stress relaxation time, and $0 < \alpha < 1$
is a parameter that quantifies the deviation of its viscoelasticity from the ideal Newtonian ($\alpha = 0$) and Maxwellian ($\alpha = 1$) behaviors. The specific values of the parameters that characterize the viscosity of these fluids are listed in Table \ref{tab:1}. The behavior of $\tilde{\eta}(s)$ as a function of the Laplace frequency for real and positive values
$s > 0$, as well as the corresponding fit curves, are plotted in Fig. \ref{fig:1}(c) for the four fluids used in the experiments. Note that the concentrations of the components of the glycerol/water mixture and those of both viscoelastic fluids were selected in such a way that their values of $\eta_0$ are close to each other. Under such conditions, in the long term, the free Brownian motion of a suspended bead would display diffusive behavior with very similar values of the diffusion coefficient given by $D = k_B T/(6\pi a \eta_0)$ for the three fluids. However, as shown in Fig. \ref{fig:1}(c), the decay of $\tilde{\eta}(s)$ with increasing $s$ is more pronounced for the micellar solution than for the polymer solution over the accessible frequency range, revealing a higher degree of viscoelasticity of the former. This is verified in Figs. \ref{fig:1}(d) and \ref{fig:1}(e), where we plot the storage and loss moduli of each fluid, which are calculated as the real and imaginary part of the complex modulus $G^*(\omega) = \mathrm{i} \omega \tilde{\eta}(s = \mathrm{i} \omega)$ using the function of Eq.~(\ref{eq:etas}) with the corresponding fitting parameters. As shown in these figures, the storage modulus of the micellar solution achieves values larger than those of the polymer solution at low frequencies, whereas the loss modulus of the former exhibits larger deviations at higher frequencies with respect to the Newtonian loss modulus of the glycerol/water mixture in comparison to that of the latter case. As demonstrated below, such subtle differences have a significant impact on the first-passage time statistics of the particle within the harmonic optical trap.  

\subsection{Determination of first-passage time events}\label{subsect:analysis}

\begin{figure}
\centering
\includegraphics[width=0.9\columnwidth]
{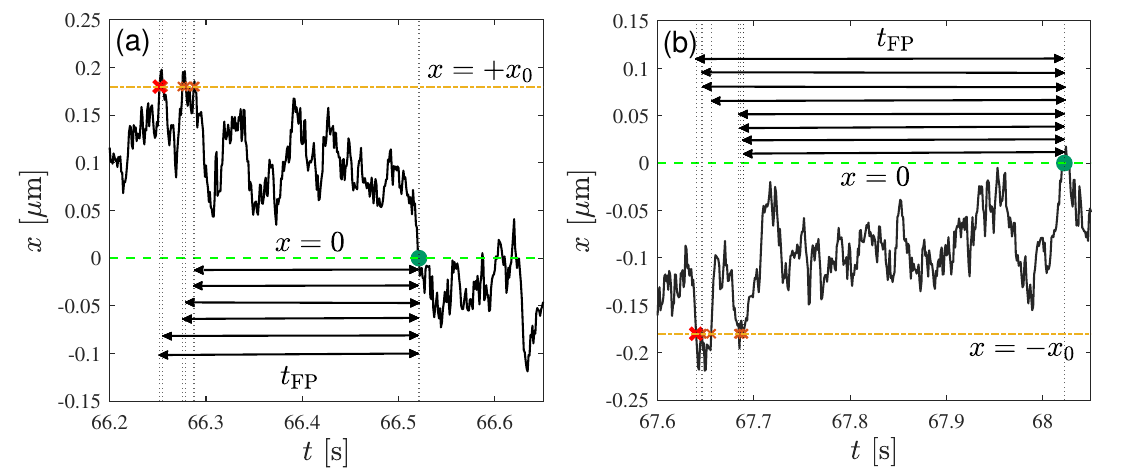}
\caption{(a) Example of the determination of first-passage events of a colloidal bead optically trapped in the micellar solution by a harmonic potential of stiffness $\kappa=1.77\times 10^{-7}~\mathrm{N~m}^{-1}$, whose durations allow to statistically sample stochastic values of $t_{\mathrm{FP}}$. Given the initial position $x_0 = 0.18\,\mu$m (dotted-dashed line) and the target point $x=0$ (dashed line), a first passage event is detected, like the trajectory segment comprised between the thick red cross (initial position) and the green circle (target). Multiple returns of the particle to the initial position (thin orange crosses) can be considered as starting points of 5 additional first-passage events provided that they eventually end at the target for the first time. (b) Example of the detection of 8 first-passage events from the initial location $-x_0 = - 0.18\,\mu$m to the target $x=0$, whose duration are also taken into account in the statistics of $t_{\mathrm{FP}}$ for the same values of $\kappa$ and $x_0$ as in Fig. \ref{fig:2}(a) due to the symmetry of the harmonic optical potential. Same symbols and line-styles as in Fig. \ref{fig:2}(a).}
\label{fig:2}
\end{figure}

Once the stiffness of a given optical potential has been determined, we measure the duration $t_{\mathrm{FP}}$
of all first-passage events performed by the bead during approximately 1 hour, starting from a predetermined initial position $x_0 \neq 0$ and reaching the target located at $x=0$ for the first time. This is exemplified in Fig. \ref{fig:2}(a), which shows an expanded view over 0.45~s of the trejectory of a particle trapped in the micellar solution by a harmonic potential of stiffness $\kappa=1.77\times 10^{-7}~\mathrm{N~m}^{-1}$, where the initial position $x_0 = +0.18\,\mu\mathrm{m} > 0$ and the target $x = 0$ are depicted as dotted-dashed and dashed lines, respectively. A first-passage event
corresponds to a portion of the trajectory that begins at $x=x_0$
and ends when reaching $x=0$
for the first time, like the segment comprised between the thick read cross (initial position) and the green circle (target), where the time elapsed between these two points corresponds to a particular value of $t_{\mathrm{FP}}$. As illustrated in Fig \ref{fig:2}(a), along this first-passage event the particle can perform multiple returns to the point $x = x_0$ before reaching the target, thus providing additional first-passage events, where their initial positions are indicated by thin orange crosses, all of them eventually ending at $x=0$, whose shorter durations must be included in the calculation of the probability density of $t_{\mathrm{FP}}$.  Note that, due to the symmetry of the potential with respect to its minimum at $x = 0$, i.e. $U(-x) = U(x)$, for a given value of $x_0 > 0$ the same analysis can be carried out for first-passage events detected between $-x_0$ and $x=0$, as illustrated in Fig. \ref{fig:2}(b), whose durations are also counted in the corresponding first-passage time statistics.

After $N$ experimental random values of the first-passage time $t_{\mathrm{FP}}$ for given values of $x_0$ and $\kappa$ have been determined along a long stochastic trajectory of the trapped bead, typically $N \approx 200 -20000$ depending on the specific values of these parameters, we calculate the probability distribution of $t_{\mathrm{FP}}$ through their normalized histogram, while the corresponding MFTP is determined by their arithmetic average. The results for various values of the parameters $x_0$ and $\kappa$ in the different fluids used in the experiments, as well as their comparison with the theoretical expressions, are shown in the next section.

\section{Results and discussion}\label{sect:res}

\subsection{Markovian baths}

\begin{figure}
\centering
\includegraphics[width=0.85\columnwidth]
{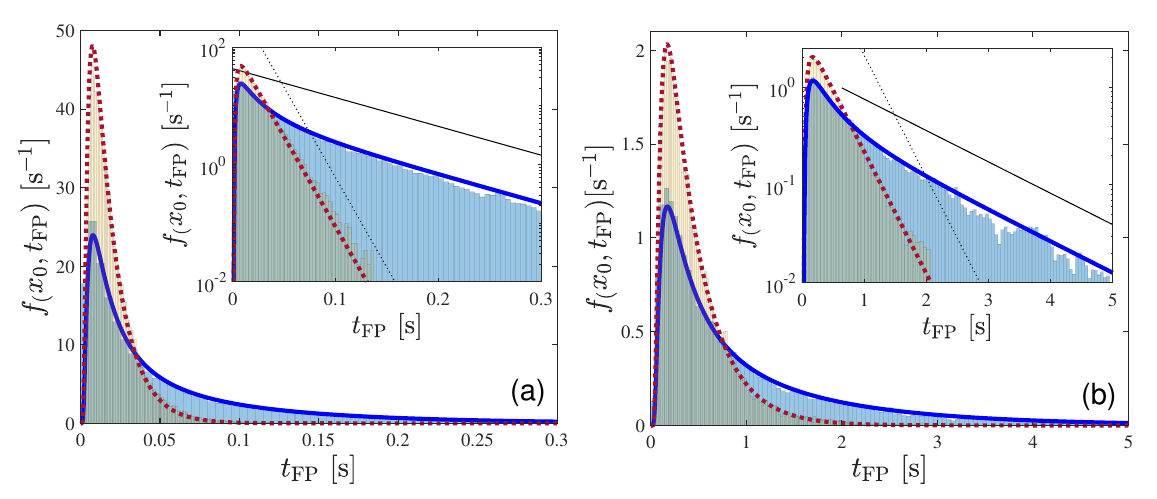}
\caption{(a) Probability density function of the first-passage time to reach the target located at $x=0$ of a particle optically trapped in water at two distinct values of the trap stiffness: $\kappa=2.13~\times~10^{-7}~\mathrm{N~m}^{-1}$ (blue bars), and $\kappa=1.33~\times~10^{-6}~\mathrm{N~m}^{-1}$ (red bars) starting from the initial position $x_0=0.1~\mu\mathrm{m}$. 
The thick solid and thick dotted lines represent the theoretical curves calculated using Eq. (\ref{eq:fptd_newton}) for these two values of $\kappa$, respectively. (b) Probability density function of the first-passage time to reach the target at $x=0$ of a particle optically trapped in the glycerol/water mixture at two distinct values of the trap stiffness: $\kappa=3.06~\times~10^{-7}~\mathrm{N~m}^{-1}$ (blue bars), and $\kappa=1.19~\times~10^{-6}~\mathrm{N~m}^{-1}$ (red bars) starting from the initial position $x_0=0.1~\mu\mathrm{m}$. The thick solid and thick dotted lines represent the theoretical curves calculated using Eq. (\ref{eq:fptd_newton}) for these two values of $\kappa$, respectively.
In both subfigures, the insets are semi-log representations
of the main plots. The thin solid and thin dotted lines in the insets depict the asymptotic behavior described by Eq. (\ref{eq:fptnewtasym}) at the corresponding small and large values of $\kappa$, respectively.}
\label{fig:3}
\end{figure}

First, we analyze the first-passage time statistics of a particle that is harmonically trapped in water, representing the reference system of this work. Even when the corresponding probability distribution given by Eq. (\ref{eq:fptd_newton}) is widely known, to the best of our knowledge, it has never been tested against experimental data. Fig. \ref{fig:3}(a) compares the experimental probability densities of $t_{\mathrm{FP}}$, $f(x_0,t_{\mathrm{FP}})$, for two different values of $\kappa$ differing by one order of magnitude ($\kappa = 2.13 \times 10^{-7}$~N~m$^{-1}$ and $\kappa = 1.33 \times 10^{-6}$~N~m$^{-1}$), in the case of the initial position $x_0 = 0.1\,\mu$m. In both cases, we verify that the theoretical formula of Eq. (\ref{eq:fptd_newton}) quantitatively describes the experimental data without fitting parameters. The probability density of $t_{\mathrm{FP}}$ vanishes as $t_{\mathrm{FP}} \rightarrow 0$, then exhibits a maximum, and converges to $f(x_0,t_{\mathrm{FP}}) = 0$ as  $t_{\mathrm{FP}} \rightarrow \infty$. The asymptotic behavior of the experimental data for large values of $t_{\mathrm{FP}}$ is in excellent agreement with Eq. (\ref{eq:fptd_newton}), which predicts the exponential decay
\begin{equation}\label{eq:fptnewtasym}
f(x_0,t_{\mathrm{FP}}) \propto e^{-\kappa t_{\mathrm{FP}}/\gamma},
\end{equation}
for $t_{\mathrm{FP}} \gg \gamma/\kappa$, as shown in the inset of Fig. \ref{fig:3}(a). Therefore, as expected, increasing the stiffness decreases the probability of observing very large values of the first-passage time.

A similar behavior of the first-passage time distribution is observed in the case of a particle trapped in the glycerol/water mixture, as demonstrated in Fig. \ref{fig:3}(b) for values of the trap stiffness similar to those used in the experiments with ultrapure water ($\kappa = 3.06 \times 10^{-7}$~N~m$^{-1}$ and $\kappa = 1.19 \times 10^{-6}$~N~m$^{-1}$), where the experimental results are also quantitatively consistent with Eq. (\ref{eq:fptd_newton}). Note that, because the viscosity of the glycerol/water mixture is approximately 20 times larger than that of ultrapure water, the resulting probability densities of the first-passage time are broader in the former case, displaying exponential tails described by Eq. (\ref{eq:fptnewtasym}) that decay correspondingly about 20 times slower than in ultrapure water for similar values of $\kappa$.

\begin{figure}
\centering
\includegraphics[width=0.9\columnwidth]
{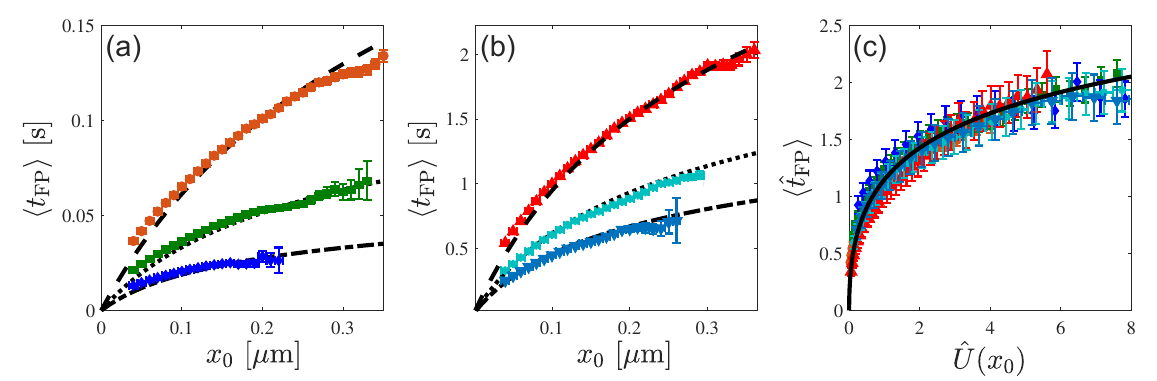}
\caption{(a) MFPT to the target located at $x=0$ as a function of the initial position $x_0$ of a particle optically trapped in water that was experimentally determined (symbols) and calculated using Eq. (\ref{eq:MFPTNewt}) (lines) at different values of the trap stiffness: $\kappa=2.13~\times~10^{-7}~\mathrm{N~m}^{-1}$ (orange $\bigcirc$, dashed line), $\kappa=5.75~\times~10^{-7}~\mathrm{N~m}^{-1}$ (green $\square$, dotted line), and $\kappa=1.33~\times~10^{-6}~\mathrm{N~m}^{-1}$ (blue $\Diamond$, dotted-dashed line). (b) MFPT to the target located at $x=0$ as a function of the initial position $x_0$ of a particle optically trapped in the glycerol/water mixture that was experimentally determined (symbols) and calculated using Eq. (\ref{eq:MFPTNewt}) (lines) at different values of the trap stiffness: $\kappa=3.06~\times~10^{-7}~\mathrm{N~m}^{-1}$ (red $\bigtriangleup$, dashed line), $\kappa=7.41~\times~10^{-7}~\mathrm{N~m}^{-1}$ (turquoise $\ast$, dotted line), and $\kappa=1.19~\times~10^{-6}~\mathrm{N~m}^{-1}$ (blue $\bigtriangledown$, dotted-dashed line). (c) Rescaled MFPT as a function of the rescaled potential energy at the initial particle position using the experimental data shown in Figs. \ref{fig:4}(a) and \ref{fig:4}(b) measured in water and the glycerol/water mixture, respectively, depicted with the same symbols and colors. The solid line represents the general expression for the rescaled MFDT in a viscous medium given by Eq. (\ref{eq:MFPTNewtadim}).}
\label{fig:4}
\end{figure}

The MFPT, $\langle t_{\mathrm{FP}} \rangle$, as a function of the initial position $x_0$ of a particle trapped in ultrapure water and in the glycerol/water mixture, are plotted in Figs. \ref{fig:4}(a) and \ref{fig:4}(b), respectively, for various values of $\kappa$. These curves show that $\langle t_{\mathrm{FP}} \rangle$ is a monotonically increasing function of $x_0$ and for a fixed value of $x_0$, it decreases with increasing $\kappa$. The experimental data can be compared with the theoretical curves obtained by calculating the first moment of $f(x_0,t_{\mathrm{FP}})$, i.e. $\langle t_{\mathrm{FP}} \rangle = \int_0^{\infty} t_{\mathrm{FP}} f(x_0,t_{\mathrm{FP}}) \,dt_{\mathrm{FP}} = \int_0^{\infty} S(x_0,t)dt$. Using Eq. (\ref{eq:survival}) along with the corresponding relaxation function of a particle trapped in a viscous bath, $\chi(t) = e^{-\kappa t/\gamma}$, the following analytical expression for the MFPT can be derived \cite{chelminiak_2024}\begin{equation}\label{eq:MFPTNewt}
    \langle t_{\mathrm{FP}} \rangle = \frac{\gamma}{\kappa} \sqrt{\frac{\pi \kappa x_0^2}{2k_B T}} {}_{1} F_1 \left( \frac{1}{2} ; \frac{3}{2} ; \frac{\kappa x_0^2}{2 k_B T}\right) - \frac{\gamma}{\kappa} \frac{\kappa x_0^2}{2k_B T} {}_{2} F_2 \left(1,1;\frac{3}{2},2;\frac{\kappa x_0^2}{2k_B T} \right),
\end{equation}
where
\begin{eqnarray}
    {}_1 F_1(a;b;z) & = & \sum_{k=0}^{\infty} \frac{(a)_k}{(b)_k} \frac{z^k}{k!} ,\nonumber\\
    {}_2 F_2(a_1,a_2;b_1,b_2;z) & = & \sum_{k=0}^{\infty} \frac{(a_1)_k (a_2)_k}{(b_1)_k (b_2)_k} \frac{z^k}{k!},
\end{eqnarray}
are the Kummer confluent hypergeometric function and the generalized hypergeometric function, respectively, with $(a)_k = \mathit{\Gamma}(a+k)/\mathit{\Gamma}(a)$ the Pochhammer symbol. In Figs. \ref{fig:4}(a) and \ref{fig:4}(b) we verify that the formula of Eq. (\ref{eq:MFPTNewt}) quantitatively describes the dependence on $x_0$ and $\kappa$ of the experimental values of $\langle t_{\mathrm{FP}} \rangle$ for the particles trapped in both viscous fluids. In addition, Eq. (\ref{eq:MFPTNewt}) reveals that, under such Markovian conditions of the medium, the MFPT can be fully expressed in terms of the viscous relaxation time of the particle in the harmonic trap $\gamma/\kappa$, the thermal energy $k_B T$, and the potential energy of the particle at the initial position $U(x_0) = \kappa x_0^2 / 2$. An important consequence of such a behavior of the MFPT of a Brownian particle harmonically trapped in a viscous bath is that, upon rescaling by the characteristic time and energy scales, $\gamma/\kappa$ and $k_B T$, the dimensionless MFPT can be recast as 
\begin{equation}\label{eq:MFPTNewtadim}
    \langle \hat{t}_{\mathrm{FP}} \rangle = \sqrt{\pi \hat{U}(x_0)} {}_1 F_1\left( \frac{1}{2};\frac{3}{2} ; \hat{U}(x_0)\right) - \hat{U}(x_0) {}_2 F_2 \left( 1,1;\frac{3}{2},2; \hat{U}(x_0)\right),
\end{equation}
where $\hat{t}_{\mathrm{FP}} = \kappa t_{\mathrm{FP}}/\gamma$ and $\hat{U}(x_0) = U(x_0)/(k_B T)$. Therefore, regardless of the viscosity of the bath and the stiffness of the trap, the rescaled MFPT must follow the unique dependence on the rescaled initial potential energy of the particle given by Eq. (\ref{eq:MFPTNewtadim}). This general behavior of $\langle \hat{t}_{\mathrm{FP}} \rangle$ is demonstrated in Fig. \ref{fig:4}(c), where all the experimental curves shown in Figs. \ref{fig:4}(a) and \ref{fig:4}(b) with different values of $\kappa$ and $\gamma$ collapse on a single curve theoretically given by Eq. (\ref{eq:MFPTNewtadim}) when plotted as a function of $\hat{U}(x_0)$.

\subsection{Non-Markovian baths}

\begin{figure}
\centering
\includegraphics[width=0.7\columnwidth]
{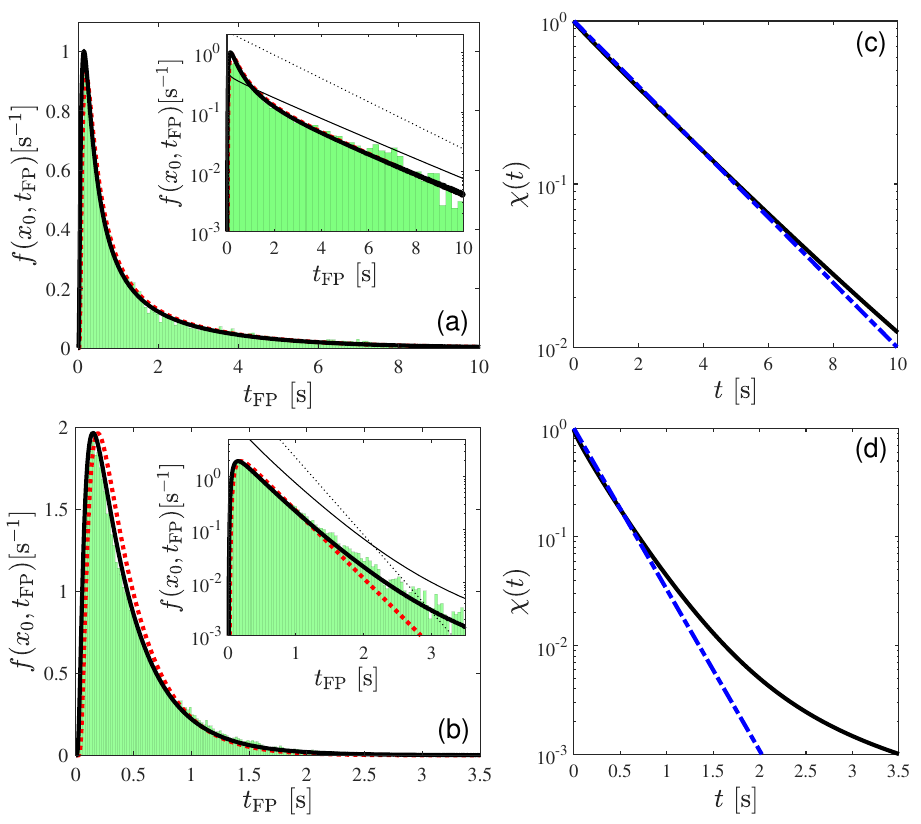}
\caption{[(a)-(b)] Probability density function of the first-passage time to the target $x=0$, $t_{\mathrm{FP}}$, of a colloidal bead optically trapped in the polymer solution, starting from $x_0=0.1~\mu\mathrm{m}$, at two different values of the trapping stiffness: (a) $ \kappa=1.88~\times~10^{-7}~\mathrm{N~m}^{-1}$, and (b) $ \kappa=1.39~\times~10^{-6}~\mathrm{N~m}^{-1}$. The experimental data are represented as vertical bars. The thick dotted and thick solid lines represent the theoretical curves computed by means of Eq. (\ref{eq:fptd_newton}) assuming a constant viscosity of the polymer solution ($\eta = \eta_0 = 0.0216$~Pa~s), and Eq. (\ref{eq:fptd}) including its frequency-dependent viscosity $\tilde{\eta}(s)$ plotted in Fig. \ref{fig:1}(c), respectively. Insets: semi-log representations of the main plots. The thin dotted and thin solid lines in the inset depict
their corresponding asymptotic behaviors described by Eqs. (\ref{eq:fptnewtasym}) and (\ref{eq:fptasymp}) respectively. 
[(c)-(d)] Time-dependence of the relaxation function $\chi(t)$ of a particle trapped in the polymer solution at: (c) $\kappa =1.88~\times~10^{-7}~\mathrm{N~m}^{-1}$, and (d) $\kappa=1.39~\times~10^{-6}~\mathrm{N~m}^{-1}$. The solid lines correspond to the correct curves that were numerically computed including the viscoelasticity of the fluid, while the dotted-dashed lines are decaying exponential functions assuming a constant viscosity equal to $\eta_0$.}
\label{fig:5}
\end{figure}

We now focus on the statistical properties of the first-passage time of a colloidal bead harmonically trapped in a non-Markovian bath with viscoelasticity. Figs. \ref{fig:5}(a) and \ref{fig:5}(b)  illustrate the main features of the distribution of the first-passage time of the particle trapped in the polymer solution to reach $x=0$ starting from $x_0 = 0.1\,\mu$m for two values of $\kappa$ that are different by one order of magnitude ($\kappa = 1.88 \times 10^{-7}$~N~m$^{-1}$ and $\kappa = 1.39 \times 10^{-6}$~N~m$^{-1}$, respectively). In both cases, we observe that the first-passage time distribution has a non-monotonic dependence on $t_{\mathrm{FP}}$, vanishing as $t_{\mathrm{FP}} \rightarrow 0$, and displaying a maximum followed by a decay to zero as $t_{\mathrm{FP}} \rightarrow \infty$, thus being qualitatively similar to that in a Markovian medium at first glance.

To understand the influence of the medium viscoelasticity on the first-passage time distribution of the particle, in Figs. \ref{fig:5}(a) and \ref{fig:5}(b) we first compare the experimental data in the polymer solution with the expression of Eq. (\ref{eq:fptd_newton}) for a Newtonian fluid using the value $\eta = \eta_0 = 0.0216$~Pa~s of the zero-shear viscosity of this viscoelastic fluid. We also show the curves calculated through the general formula of Eq. (\ref{eq:fptd}) including the viscoelastic behavior of the polymer solution. In the latter case, the relaxation function $\chi(t)$, which is plotted for each value of $\kappa$ in Figs. \ref{fig:5}(c) and \ref{fig:5}(d), is computed by Laplace inversion of Eq. (\ref{eq:chi}) using the Talbot method~\cite{talbot1979}. Here, the function of Eq. (\ref{eq:etas}) for the frequency-dependent viscosity with the corresponding fitting parameters listed in Table \ref{tab:1} is employed to express the Laplace transform of the friction memory kernel, $\tilde{\Gamma}(s) = 6\pi a \tilde{\eta}(s)$, in an analytical form. As can be seen in Figs. \ref{fig:5}(a) and \ref{fig:5}(b), the theoretical expressions of Eq. (\ref{eq:fptd_newton}) for a Newtonian fluid, and Eq. (\ref{eq:fptd}) including the effect of viscoelasticity, are very close to each other and both seem to describe well the experimental distributions. However, a close inspection of these curves in the insets of Figs. \ref{fig:5}(a) and \ref{fig:5}(b) reveals that the fluid viscoelasticity gives rise to asymptotic behaviors of the first-passage time distribution that can deviate from the exponential decay in a Newtonian fluid depending on the value of the trap stiffness. For example, in the case of the weakest optical trap ($\kappa = 1.88 \times 10^{-7}$~N~m$^{-1}$), we notice that this deviation is very small for the experimentally accessible values of $t_{\mathrm{FP}}$, and therefore the  first-passage time distribution looks almost identical to that in a purely viscous bath, see Fig. \ref{fig:5}(a). Nevertheless, when the harmonic confinement is one order of magnitude stronger, a non-exponential tail of the first passage time distribution in the polymer solution becomes conspicuous, as observed in Fig. \ref{fig:5}(b) for the experimentally determined distribution at $\kappa = 1.39 \times 10^{-6}$~N~m$^{-1}$, which is quantitatively described only by the general formula of Eq. (\ref{eq:fptd}). It should be pointed out that this non-exponential tail in the viscoelastic polymer solution discloses a slower decay of the curve $f(x_0,t_{\mathrm{FP}})$ with increasing $t_{\mathrm{FP}}$ in comparison with the corresponding asymptotic behavior in a Newtonian fluid.

Such asymptotic features of the first-passage time distribution in a viscoelastic fluid can be traced back to the behavior of the relaxation function $\chi(t)$, since Eq. (\ref{eq:fptd}) shows that, in the general case of an arbitrary friction memory kernel of the particle, for sufficiently large $t_{\mathrm{FP}}$ the first-passage time distribution has the form 
\begin{equation}\label{eq:fptasymp}
f(x_0,t_{\mathrm{FP}}) \propto -\frac{d \chi(t_{\mathrm{FP}})}{dt_{\mathrm{FP}}}.
\end{equation}
In a Newtonian fluid, this simply corresponds to the exponential decay expressed by Eq. (\ref{eq:fptnewtasym}) with a single characteristic time-scale set by $\gamma / \kappa$. In contrast, in a viscoelastic fluid, the relaxation function exhibits a more complex behavior due to the interplay between the distinct time-scales of the system that emerge from the coupling of the particle with its non-Markovian environment. Indeed, Fig. \ref{fig:5}(c) shows that for small $\kappa$, $\chi(t)$ has a quasi-exponential behavior, and therefore $f(x_0,t_{\mathrm{FP}})$ is well approximated by Eq. (\ref{eq:fptd_newton}) with constant friction coefficient $\gamma = 6\pi a \eta_0$, where $-\frac{d\chi(t_{\mathrm{FP}})}{dt_{\mathrm{FP}}} \approx \frac{\kappa}{\gamma} e^{-\kappa t_{\mathrm{FP}} / \gamma}$. At higher $\kappa$, the non-exponential behavior of $\chi(t)$ becomes manifest, see Fig. \ref{fig:5}(d), thus leading to a non-exponential behavior of $-\frac{d\chi(t_{\mathrm{FP}})}{dt_{\mathrm{FP}}}$ at large values of $t_{\mathrm{FP}}$ that explains the tail of $f(x_0,t_{\mathrm{FP}})$ decaying more slowly than $e^{-\kappa t_{\mathrm{FP}}/\gamma}$.

\begin{figure}
\centering
\includegraphics[width=0.7\columnwidth]
{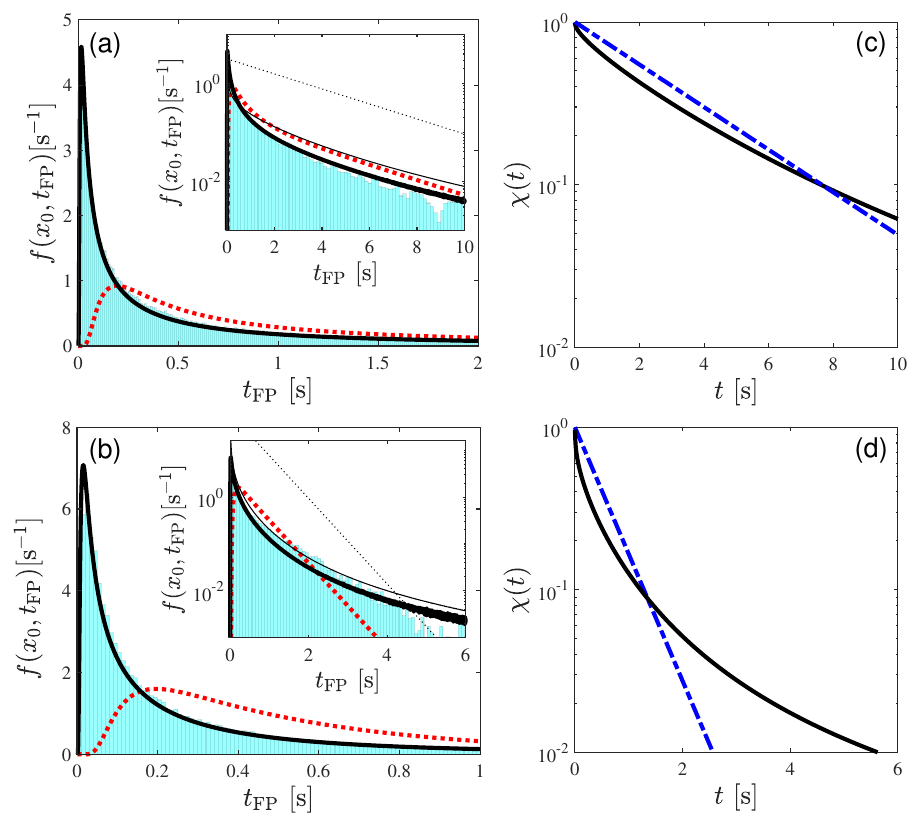}
\caption{[(a)-(b)] Probability density function of the first-passage time to the target $x=0$, $t_{\mathrm{FP}}$, of a colloidal bead optically trapped in the micellar solution, starting from $x_0=0.1~\mu\mathrm{m}$, at two different values of the trapping stiffness: (a) $ \kappa=1.77~\times~10^{-7}~\mathrm{N~m}^{-1}$, and (b) $ \kappa=9.89~\times~10^{-7}~\mathrm{N~m}^{-1}$. The experimental data are represented as vertical bars. The thick dotted and thick solid lines represent the theoretical curves computed by means of Eq. (\ref{eq:fptd_newton}) assuming a constant viscosity of the micellar solution ($\eta = \eta_0 = 0.0292$~Pa~s), and Eq. (\ref{eq:fptd}) including its frequency-dependent viscosity $\tilde{\eta}(s)$ plotted in Fig. \ref{fig:1}(c), respectively. Insets: semi-log representations of the main plots. The thin dotted and thin solid lines in the insets depict
their corresponding asymptotic behaviors described by Eqs. (\ref{eq:fptnewtasym}) and (\ref{eq:fptasymp}) respectively. 
[(c)-(d)] Time-dependence of the relaxation function $\chi(t)$ of a particle trapped in the micellar solution at: (c) $\kappa =1.77~\times~10^{-7}~\mathrm{N~m}^{-1}$ (c), and (d) $\kappa=9.89~\times~10^{-7}~\mathrm{N~m}^{-1}$. The solid lines correspond to the correct curves that were numerically computed including the viscoelasticity of the fluid, while the dotted-dashed lines are decaying exponential functions assuming a constant viscosity equal to $\eta_0$.}
\label{fig:6}
\end{figure}

The above-mentioned behavior of the first-passage time distribution of a particle trapped in a non-Markovian bath is more pronounced in the case of the micellar solution, which exhibits a higher degree of viscoelasticity than the polymer solution, as shown in subsection \ref{subsect:microrheo}. This is demonstrated in Figs \ref{fig:6}(a) and \ref{fig:6}(b), which shows the experimentally determined probabilities densities of $t_{\mathrm{FP}}$ for a particle trapped in the micellar solution by harmonic potentials of stiffness $\kappa = 1.77 \times 10^{-7}$~N~m$^{-1}$ and $\kappa = 9.89 \times 10^{-7}$~N~m$^{-1}$, respectively. The theoretical distributions calculated according to Eqs. (\ref{eq:fptd_newton}) assuming a Newtonian fluid of viscosity $\eta = \eta_0 = 0.0292$~Pa~s, i.e. equal to the zero-shear viscosity of the micellar solution, and (\ref{eq:fptd}) including the frequency-dependent viscosity of the micellar solution in the relaxation function, are also plotted for comparison. In this viscoelastic fluid, there is a significant difference between these two theoretical curves for both values of $\kappa$, where that obtained using Eq. (\ref{eq:fptd_newton}) is unable to capture the behavior of the experimental distribution, as it predicts a maximum at a value of $t_{\mathrm{FP}}$ one order of magnitude larger, followed by an exponential tail that decreases faster than the experimental data. By contrast, for both values of $\kappa$ the theoretical curve that includes the viscoelastic response of the micellar solution describes remarkably well the full behavior of the experimental first-passage time distribution over all accessible values of $t_{\mathrm{FP}}$. In particular, the location of the maximum and the non-exponential tail derived from Eq. (\ref{eq:fptasymp}) are both reproduced by Eq. (\ref{eq:fptd}). It should also be noticed that, similarly to the first-passage time of a particle in the polymer solution, the deviations of the corresponding probability density with respect to the behavior in a Newtonian fluid of the same zero-shear viscosity become more pronounced in the viscoelastic micellar solution with increasing trap stiffness due to long-lived temporal correlations induced by the harmonic confinement. This can be confirmed in Figs \ref{fig:6}(c) and \ref{fig:6}(d), which shows that the non-exponential behavior of the relaxation function $\chi(t)$ is much more prominent for $\kappa = 9.89 \times 10^{-7}$~N~m$^{-1}$ than for $\kappa = 1.77 \times 10^{-7}$~N~m$^{-1}$.    

\begin{figure}
\centering
\includegraphics[width=0.7\columnwidth]
{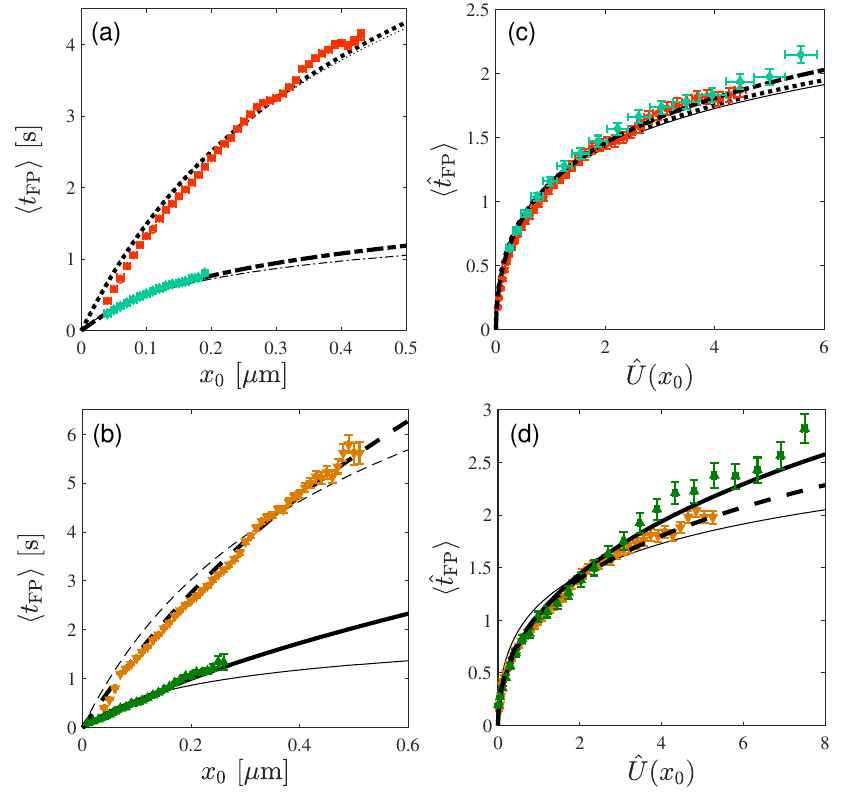}
\caption{[(a)-(b)] Dependence of the MFPT to the target position ($x=0$) on the initial location $x_0$, experimentally measured (symbols) and numerically computed (thick lines) of a particle harmonically trapped in: (a) the polymer solution at $\kappa=1.88~\times~10^{-7}~\mathrm{N~m}^{-1}$ (red $\bigcirc$, thick dotted line), and $\kappa=1.39~\times~10^{-6}~\mathrm{N~m}^{-1}$ (teal $\Diamond$, thick dotted-dashed line); and (b) the micellar solution at $\kappa=1.77~\times~10^{-7}~\mathrm{N~m}^{-1}$ (orange $\bigtriangledown$, thick dashed line) and $\kappa=9.89~\times~10^{-7}~\mathrm{N~m}^{-1}$ (green $\bigtriangleup$, thick solid line). In both subfigures, thin lines depict the MFPT curves described by Eq. (\ref{eq:MFPTNewt}) assuming frequency-independent viscosities $\eta = \eta_0 = 0.0216$~Pa~s and $\eta = \eta_0 = 0.0292$~Pa~s of the polymer and the micellar solution, respectively, whose line style portrays the same value of $\kappa$ used to determine the corresponding thick curve including the viscoelasticity of the medium. [(c)-(d)] Rescaled
MFPT of the trapped particle as a function of its rescaled initial potential energy using the experimental data shown in Figs. \ref{fig:7}(a) and \ref{fig:7}(b) in the case of: (c) the polymer solution, and (d) the micellar solution, respectively. Same symbols and line styles as in Figs. \ref{fig:7}(a) and \ref{fig:7}(b). In both subfigures, the thin solid line represents the general expression of Eq. (\ref{eq:MFPTNewtadim}) for the rescaled MFPT of a particle trapped in a purely viscous medium.}
\label{fig:7}
\end{figure}

Finally, we examine the impact of the fluid viscoelasticity on the MFPT of the trapped particle. This is exemplified in Figs. \ref{fig:7}(a) and \ref{fig:7}(b), which show the dependence of $\langle t_{\mathrm{FP}} \rangle$
on its initial location $x_0$ determined experimentally for two different values of the trap stiffness in the polymer and the micellar solution, respectively. In these graphs, we also contrast the theoretical curves calculated in a Newtonian fluid using Eq. (\ref{eq:MFPTNewt}), i.e. neglecting viscoelasticity, and by the numerical calculation of the integral $\langle t_{\mathrm{FP}} \rangle = \int_0^{\infty} S(x_0,t)\,dt$, where we substitute the numerically obtained relaxation curve $\chi(t)$ including viscoelasticity in the expression of the survival probability given by Eq. (\ref{eq:survival}). As expected from the observed properties of $f(x_0,t_{\mathrm{FP}})$, in all cases we find a very good agreement between the experimental MFPT curves and the numerical ones including the viscoelasticity of the medium. Moreover, in the case of the particle motion in the polymer solution at low trap stiffness ($\kappa = 1.88 \times 10^{-7}$~N~m$^{-1}$), the difference of the experimental MFPT curve from the theoretical one corresponding to a Markovian viscous bath, Eq. (\ref{eq:MFPTNewt}), is very small. However, a tighter confinement or an increasing viscoelasticity of the medium clearly reveal a systematic deviation of the actual MFPT curve from that obtained assuming a viscous bath, as confirmed in Fig. \ref{fig:7}(a) in the case of the polymer solution at larger trap stiffness ($\kappa = 1.39 \times 10^{-6}$~N~m$^{-1}$), and in Fig. \ref{fig:7}(b) for the micellar solution at all explored values of $\kappa$. These deviations consist of shorter MFPT values at sufficiently small initial distances $x_0$ followed by a crossover at farther positions $x_0$ to an augmented MFPT relative to the values estimated in a Newtonian environment of the same zero-sehar viscosity. 

\begin{figure}
\centering
\includegraphics[width=0.7\columnwidth]
{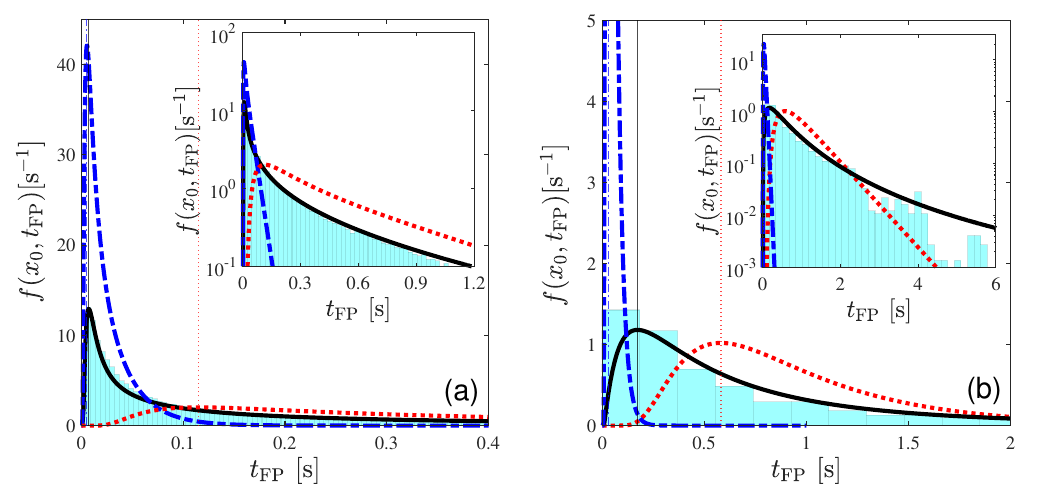}
\caption{First-passage time distribution to the target located at $x=0$ of a colloidal bead trapped in the micellar solution by an optical potential of stiffness $\kappa = 9.89\times10^{-7}~\mathrm{N~m}^{-1}$ for two distinct initial positions $x_0$, corresponding to two different values of its rescaled potential energy $\hat{U}(x_0)$: (a) $x_0 = 0.07\,\mu$m, $\hat{U}(x_0) = 0.59$; (b) $x_0 = 0.2\,\mu$m, $\hat{U}(x_0) = 4.86$. In both subfigures, vertical bars represent the experimental data, solid lines are theoretical curves that were calculated through Eq. (\ref{eq:fptd}) including the viscoelasticity of the medium, whereas curves depicted by thick broken lines were determined by means of Eq. (\ref{eq:fptd_newton}) assuming constant values of the fluid viscosity: $\eta = \eta_0 = 0.0292$~Pa~s (dotted lines), and $\eta = \eta_{\infty} = 0.0014$~Pa~s (dotted-dashed lines). Thin vertical lines depict the location of the maxima of the curves traced with the same line style. Insets are semi-log representations of the main figures.}
\label{fig:8}
\end{figure}

Motivated by the general rescaled form of the MFDT of a particle trapped in a Newtonian environment described by Eq. (\ref{eq:MFPTNewtadim}), we perform a similar rescaling in the case of the polymer and micellar solutions, where we use their zero-shear viscosity $\eta_0$ in the calculation of the corresponding friction coefficient $\gamma = 6 \pi a \eta_0$ to determine the characteristic relaxation time $\gamma/\kappa$. In Figs. \ref{fig:7}(c) and \ref{fig:7}(d) we show the results of the rescaled MFPT, $\langle \hat{t}_{\mathrm{FP}} \rangle = \kappa \langle t_{\mathrm{FP}} \rangle / \gamma$, as a function of the dimensionless potential energy at the initial position, $\hat{U}(x_0) = U(x_0)/(k_B T)$. These plots clearly mirror the main features of the MFPT that were previously described, namely, shorter rescaled MFPT values at sufficiently small initial potential energy, typically $\hat{U}(x_0) \lesssim 1$, and increasingly larger values compared to the rescaled MFPT curve occurring in a Newtonian bath. Furthermore, in stark contrast to the universal behavior in a viscous fluid illustrated in Fig. \ref{fig:4}(c), the rescaled MFPT curve in viscoelastic media exhibits a dependence on the stiffness of the trap that confines the particle. This dependence can be attributed to the non-exponential temporal correlations induced by the different time-scales of the system and the bath with increasing trap stiffness.
Indeed, Fig. \ref{fig:8}(a) shows that, for a strong trap stiffness ($\kappa = 9.89\times10^{-7}~\mathrm{N~m}^{-1}$) but sufficiently small initial position ($x_0 = 0.07\,\mu$m) such that the corresponding rescaled potential energy of the particle in the micellar solution is $\hat{U}(x_0) = 0.59 < 1$, $f(t_{\mathrm{FP}},x_0)$ exhibits a sharp peak, followed by a smoother decay. The location of the maximum of the peak ($\langle t_{\mathrm{FP}}\rangle = 0.007$~s) is much closer to the maximum that would be observed in a Newtonian fluid of viscosity equal to $\eta = \eta_{\infty} = 0.0014$~Pa~s ($\langle t_{\mathrm{FP}}\rangle = 0.006$~s) than that in a viscous one with $\eta = \eta_0 = 0.0292$~Pa~s ($\langle t_{\mathrm{FP}}\rangle = 0.115$~s). This shows that for small values of $\hat{U}(x_0)$, very fast first-passage events that are mainly driven by the diffusive motion of the particle through the solvent with viscosity $\eta_{\infty}$ dominate over the slower ones that result from the relaxation of the viscoelastic microstructure of the medium, thus giving rise to the apparent reduction of $\langle \hat{t}_{\mathrm{FP}} \rangle$ relative to that described by Eq. (\ref{eq:MFPTNewtadim}). On the other hand, Fig. \ref{fig:8}(b) reveals that, for the same strong trap stiffness ($\kappa = 9.89\times10^{-7}~\mathrm{N~m}^{-1}$) and large initial distances from the target ($x_0 = 0.2\,\mu$m) such that $\hat{U}(x_0) = 4.86 > 1$, the non-exponential tail of the first passage-time distribution becomes much more prominent, while its maximum is shifted to a value ($\langle t_{\mathrm{FP}}\rangle = 0.171$~s) that is closer to that of the corresponding first-passage time distribution of a particle in a Newtonian fluid of viscosity $\eta_0$ ($\langle t_{\mathrm{FP}}\rangle = 0.581$~s) rather than the maximum in a fluid with viscosity $\eta_{\infty}$ ($\langle t_{\mathrm{FP}}\rangle = 0.029$~s). Therefore, long-lived first-passage events, which are absent in a purely viscous medium, lead to a systematic growth of the average time it takes for the particle to reach the origin starting with a sufficiently high potential energy.

\section{Summary and concluding remarks}\label{sect:summ}

In this work, we have 
analyzed the statistics of the time it takes for an overdamped Brownian particle confined in a fluid by a harmonic potential to reach the minimum for the first time. 
Using the generalized Langevin equation, we have derived a general expression for the probability density function of this first-passage time, which explicitly includes a  function that describes both the relaxation of the particle position from a given initial position and its stationary autocorrelation function. We have checked that the derived formula reduces to a well-known expression for this first-passage process in a viscous bath. Our results were experimentally tested using colloidal beads optically trapped in diverse fluids, displaying first-passage time distributions with asymptotic exponential decays in viscous (Newtonian) fluids, and slower non-exponential tails in viscoelastic ones. We have verified that, while the mean first-passage time of a particle confined in a Newtonian fluid exhibits a general rescaled form, such a rescaling is not possible in a viscoelastic fluid, since this average quantity depends on the trap stiffness due to the coupling between the time-scales of the system and those of the bath. In particular, the mean first-passage time of the particle kept at high stiffness in a viscoelastic environment can become shorter or longer than that in a viscous fluid of the same zero-shear viscosity depending on the initial particle position. When the potential energy of the particle at its starting point is smaller than the thermal energy, its first-passage events are dominated by diffusion in the solvent, thus shortening their duration on average, whereas the slowly-decaying correlations that emerge at large trap stiffness increase the mean first-passage time if the initial potential energy is larger than the thermal one.

Recently, there has been a growing interest in investigating first-passage processes under non-Markovian conditions, as they frequently occur in many small-scaled systems where  purely diffusive models fail to adequately describe their stochastic dynamics, such as Brownian particles moving in complex fluid environments. The results presented in this paper provide insight into the impact of viscoelasticity of the medium on the first-passage time statistics of a harmonically confined colloidal bead, which represents a minimal setup where non-Markovian effects can be clearly identified. Further extensions of this work could be carried out for the first-passage-time statistics of other confined Brownian systems, such as passive particles in double-well potentials \cite{ferrer_2024}, and optically trapped active particles \cite{darabi2026}, either under constant or intermittent conditions.

\section*{Author Contributions}
{\bf{B. R. Ferrer}}: conceptualization, data curation, formal analysis, investigation, methodology,
software, validation, visualization. {\bf{J. R. Gomez-Solano}}: conceptualization, data curation,
formal analysis, funding acquisition, investigation, methodology, project administration,
resources, software, supervision, validation, visualization, writing – original draft.

\section*{Acknowledgments}

We acknowledge the support from DGAPA-UNAM PAPIIT Grant No. IN110324 and PIIF-2-24.

\end{document}